\documentclass[headings,letterpaper,twoside,twocolumn,letterpaper]{elsart3p}

\usepackage{amssymb}
\usepackage{url}
\usepackage{epsfig}
\usepackage{graphicx}
\usepackage{lineno}

\usepackage{ifpdf}

\newif\ifpdf
\ifx\pdfoutput\undefined
   \pdffalse               
\else
   \pdfoutput=1            
   \pdftrue
\fi

\ifpdf
\usepackage[%
  pdftitle={The use of Geant4 for simulations of a plastic
       $\beta$-detector and its application to efficiency calibration},
  pdfauthor={V.V.~Golovko, V. E. Iacob and J. C. Hardy},
  pdfsubject={Geant4 for simulations of a plastic beta-detector efficiency},
  pdfkeywords={Precise $\beta$-branching-ratio measurements; the Standard Model;
               unitarity of the Cabibbo-Kobayashi-Moskava matrix; Monte Carlo
               calculations; Geant4 efficiency calculation for plastic
               beta-detector.},
  pdfstartview=FitH,
  bookmarks=true,
  bookmarksopen=true,
  breaklinks=true,
  colorlinks=true,
  linkcolor=blue,anchorcolor=blue,
  citecolor=blue,filecolor=blue,
  menucolor=blue,pagecolor=blue,
  urlcolor=blue]{hyperref}
\else
\usepackage[%
  breaklinks=true,%
  colorlinks=true,%
  linkcolor=blue,anchorcolor=blue,%
  citecolor=blue,filecolor=blue,%
  menucolor=blue,pagecolor=blue,%
  urlcolor=blue]{hyperref}
\fi

\runtitle{Geant4 for simulations of a $\beta$-detector efficiency}
\runauthor{V.V.~Golovko \textit{et. al}.}

\begin{document}

\begin{frontmatter}
\title{The use of Geant4 for simulations of a plastic
       $\beta$-detector and its application to efficiency calibration}

\author[label1]{V.V.~Golovko\thanksref{label2}}
\ead{vgolovko@comp.tamu.edu}
\author[label1]{V. E. Iacob}
\author[label1]{J. C. Hardy}
\thanks[label2]{This work was supported by the U.S. Department of
                 Energy under Grant No. DE-FG03-93ER40773 and by the Robert A. Welch
                 Foundation under Grant No. A-1397.}
\address[label1]{Cyclotron Institute, Texas
          A$\mathbf{\&}$M University, College Station, TX 77843-3366,
          USA}

\begin{abstract}
Precise
$\beta$-branching-ratio measurements are required in order to
determine \textit{ft}-values as 
part of our program to test the Electroweak Standard Model via
unitarity of the Cabibbo-Kobayashi-Moskawa matrix. For the
measurements to be useful in this test, their precision must be
close to 0.1{\%}. In a branching-ratio measurement, we position the
radioactive sample between a thin plastic scintillator used to
detect $\beta $-particles, and a HPGe detector for $\gamma $-rays.
Both $\beta$ singles and $\beta$-$\gamma$ coincidences are recorded.
Although the branching ratio depends most strongly on the HPGe
detector efficiency, it has some sensitivity to the energy dependence
of the $\beta $-detector efficiency.  We report here on a study of
our $\beta $-detector response function, which used Monte Carlo
calculations performed by the Geant4 toolkit.  Results of the
simulations are compared to measured $\beta $-spectra from several
standard $\beta$-sources.
\end{abstract}

\end{frontmatter}


\section{Introduction}

Knowledge of the total efficiency
of a plastic $\beta$-detector is crucial to our precision
experiments testing the Electroweak Standard Model. We measure the
$ft$ values for superallowed $0^+$~$\rightarrow$~$0^+$ nuclear
transitions, from which we obtain the value of V$_{ud}$, the up-down
quark-mixing element of the Cabibbo-Kobayashi-Moskawa (CKM) matrix.
This requires that half-lives, branching ratios and decay energies
all be measured with high precision, 0.1\% or better.  (The most
recent complete review of this work can be found in Ref.
\cite{hardy:055501}, with an update in Ref. \cite{Towner:025501}). Since
branching ratios are typically determined from the intensities of
$\beta$-delayed $\gamma$ rays that are observed in $\beta$-$\gamma$
coincidence measurements, it is the $\gamma$-ray detector's
efficiency that is most crucial in aiming for 0.1\% precision.
However, a good knowledge of the energy-dependence of the
$\beta$-detector's efficiency is also required. Here we report
studies of our $\beta$-detector's response function, with source
measurements and Monte Carlo calculations performed with the Geant4
(version~4.9.0) toolkit~\cite{agostinelli:03}.

In a typical measurement of a $\beta$-decay branching ratio (see,
for example, Ref.~\cite{iacob:015501,Hardy2003}), we implant a
radioactive species into Mylar tape, then rapidly move the tape to a
shielded counting station, where the sample is positioned between a
1-mm-thick plastic scintillator to detect $\beta$-particles, and a
70\% HPGe detector to detect $\gamma$-rays.  We record both $\beta$
singles and $\beta$-$\gamma$ coincident events.  Since the
efficiency of the HPGe detector has been very precisely
determined~\cite{Helmer:2003}, to first order a $\beta$-branching
ratio is simply given by the measured number of coincident $\gamma$
rays that follow that $\beta$ branch divided by the total number of
$\beta$ singles from all branches; the $\beta$-detector efficiency
simply cancels out and need not be determined.  For high precision,
though, it becomes necessary to account for the slightly different
efficiency of the $\beta$ detector for each $\beta$ transition, a
difference that affects the measured intensities of the coincident
$\gamma$ rays.

\begin{figure}[!]
   \includegraphics[width=\columnwidth]{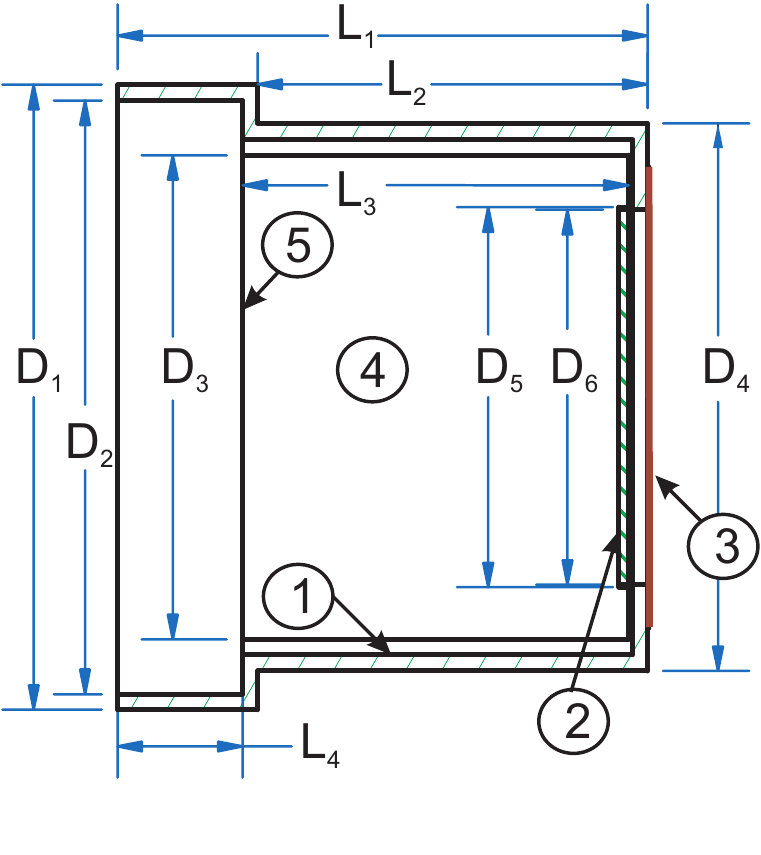}\\[-40pt]
   \caption{Schematic drawing of the $\beta$-detector. Labeled dimensions are
  given in Table~\ref{tab:dimensions}; others appear in the text.
  Incident $\beta$-particles from a radioactive source
  pass through a thin havar-foil window (3) and into the plastic-scintillator
  disc (2). The disc is recessed into a lucite light guide (4), which is
  coupled to a phototube (not shown in the figure) at surface 5. The scintillator
  and light guide are enclosed in a cylindrical cover (1) made from PVC.}
  \label{fig:betaDet}
\end{figure}

The energy dependence of our $\beta$-detection efficiency is caused
principally by the low-energy electronic threshold, which removes a
slightly different fraction of the total $\beta$ spectrum for
different end-point energies.  Since our threshold is at $\sim$80
keV and, for our superallowed-decay studies, end-point energies are
typically 2 MeV or more, we lose at most a few percent of the total
$\beta$ particles for any single transition.  Thus the change in
this loss from transition to transition in the same decay is even
smaller.  Nevertheless, the precision we strive for in our
branching-ratio measurements is very high and we seek to account
reliably for that energy dependence. Even though a Monte Carlo code
like Geant4 should be well suited to simulating the response
function of a thin plastic scintillator, we considered it important
to test and evaluate the code's results first, by comparing them
against experimental data taken with several $\beta$ sources that
also emit conversion-electrons: $^{133}$Ba, $^{137}$Cs and
$^{207}$Bi.

The Geant4 toolkit used in treating the transportation of $\beta$ and $\gamma$
particles through matter is both modular and flexible, especially in
the description of low-energy electromagnetic processes down to
250~eV (see, for example Ref.~\cite{apostolakis:99,Chauvie:2004}).
In addition, it is also possible to simulate rather complicated 3-D
geometry, select a variety of materials and decay products
(including radioactive ions), and choose how to handle the physical
processes governing particle interactions. Moreover, it provides
output of the simulated data at different stages in the calculation
and under various selection criteria.  An overview of recent
developments in diverse areas of this toolkit is presented in
Ref.~\cite{Allison:2006}.

\begin{table}[b]
  \vspace{10pt}
  \centering
  \caption{Measured detector dimensions used in our Monte Carlo
   calculations. Letters in the second column correspond to
   labels in Fig.~\ref{fig:betaDet}.}
   \label{tab:dimensions}
   \begin{tabular*}{\columnwidth}{@{\extracolsep{\fill}}lcc}\\[-3mm]
     \hline
     \hline
           \begin{tabular}{l}
          Detector parameter \\
      \end{tabular}
     & &
      \begin{tabular}{c}
          Value (mm) \\
      \end{tabular}
    \\
     \hline
     Outer shoulder, {\o}    & D$_1$ & 63.50 \\[-1mm]
     Inner shoulder, {\o}    & D$_2$ & 60.33 \\[-1mm]
     Light guide, {\o}       & D$_3$ & 49.20 \\[-1mm]
     Outer PVC cover, {\o}   & D$_4$ & 55.63 \\[-1mm]
     Hole in PVC cover, {\o}  & D$_5$ & 38.61 \\[-1mm]
     Plastic scintillator, {\o}  & D$_6$ & 38.10 \\[-1mm]
     Length of PVC cover     & L$_1$ & 53.85 \\[-1mm]
     Length of outer shoulder& L$_2$ & 39.62 \\[-1mm]
     Length of light guide   & L$_3$ & 39.24 \\[-1mm]
     Length of inner shoulder& L$_4$ & 12.70 \\
     \hline
     \hline
   \end{tabular*}
\end{table}

\section{\label{Det}Detector Arrangement and Measurements}

The $\beta$-detector assembly is illustrated in
Fig.~\ref{fig:betaDet}, with detailed dimensions given in
Table~\ref{tab:dimensions}.  It consists of a 1-mm-thick Bicron
BC404 scintillator disc recessed into a cylindrical Lucite light
guide, to which it is optically coupled.  The light guide, in turn,
is optically coupled to a photomultiplier tube (not shown in the
figure and not included in the simulations). Optical cement (BC-600)
from Bicron was applied to the surfaces between the scintillator and
the light guide, and between the light guide and the
photomultiplier tube (R329P from Hamamatsu). The scintillator disc,
light guide and the last 13 mm of the phototube are enclosed in an
opaque cylindrical shield made from 1.5-mm-thick polyvinyl chloride
(PVC). The opening in the scintillator end of the PVC shield is
slightly larger in diameter than the scintillator disc, and is
covered with a pin-hole-free, 5-$\mu$m-thick havar foil. The
$\beta$-particles enter the detector assembly through this foil with
essentially negligible energy loss.

\begin{figure}[t]
  \includegraphics[width=\columnwidth]{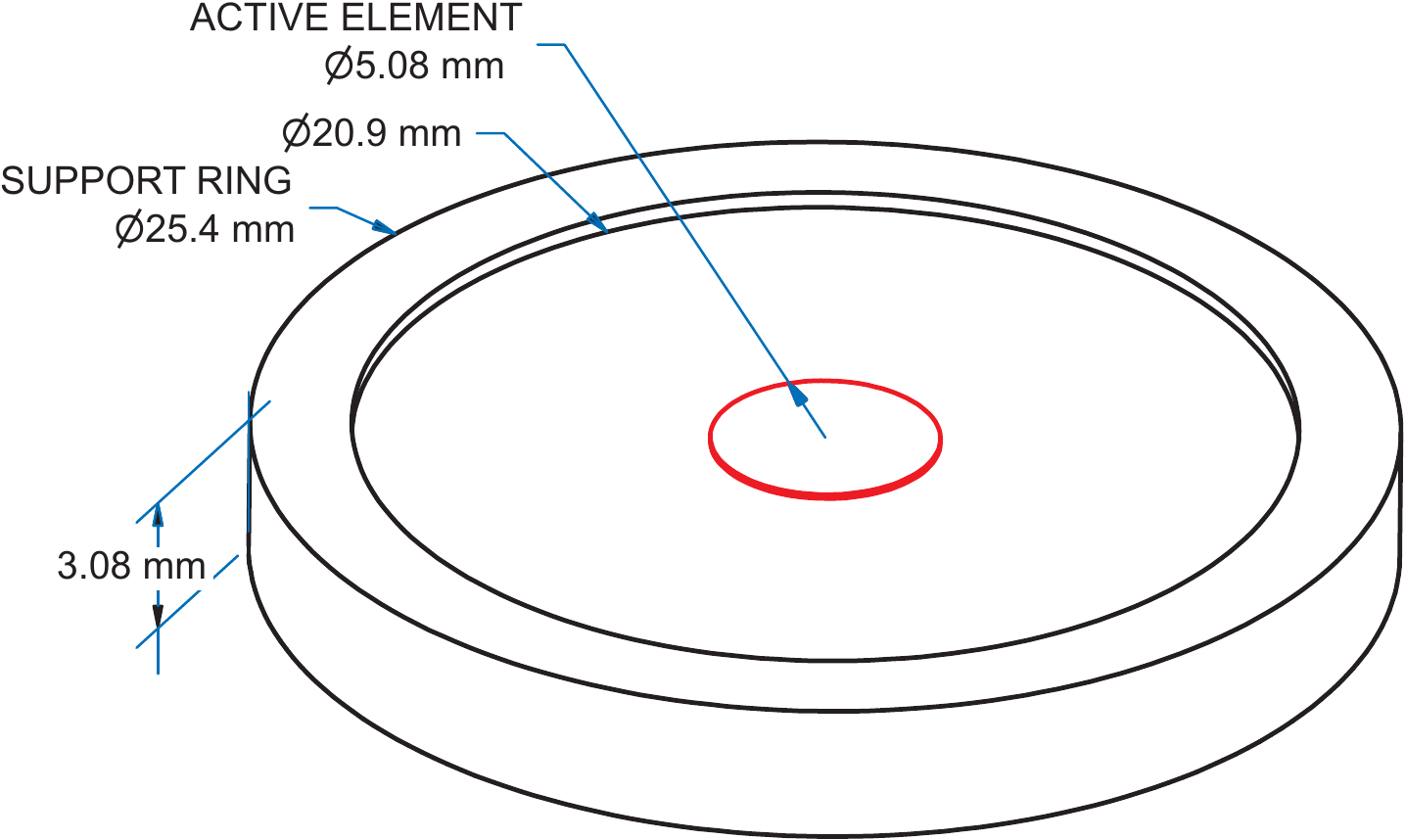}\\[-10pt]
  \caption{Radioactive-source geometry: the support ring is made from aluminum.}
  \label{RadioactiveSource2}
\end{figure}

For our measurements of the detector response function, each
radioactive source was placed at a distance of 13~mm from the havar
window of the detector assembly and was axially aligned with it.  The
distance was determined with the aid of an AccuRange 600$^{\rm{TM}}$ Laser
Displacement Sensor (model AR600-4000)~\cite{LaserSensor_web}, which measures
distance with an absolute precision better than 0.1~mm.  Both detector and source
were placed on stands on a lab table (in air at atmospheric
pressure) as far away as possible from other objects. We used three
different 37-kBq sources -- $^{133}$Ba, $^{137}$Cs and $^{207}$Bi.
All three were open sources sold by Isotope Products Laboratories as
``conversion electron sources."  Each source, being specially
prepared to minimize scattering or degradation of the emitted
electrons, had been deposited as a 5-mm-diameter spot on a thin foil
-- stainless steel with a thickness of 51~$\mu$m in the case of
$^{207}$Bi, aluminized Mylar with a thickness of 6~$\mu$m for the
other two -- and covered only by a 100-$\mu$g/cm$^2$ acrylic film.
The source-holder geometry is shown in
Fig.~\ref{RadioactiveSource2}.

All three of these radioactive sources emit $\gamma$ rays as well as
electrons and, although our thin scintillator is relatively
insensitive to the former, we nonetheless took extra precautions to
ensure that we were only studying the detector's response to the
latter. In addition to recording a spectrum from each source as
already described, we also recorded a second spectrum with a
2-mm-thick aluminum plate inserted between the source and the
detector.  This plate was thick enough to remove all the $\beta$
particles without significantly attenuating the $\gamma$ rays. We
then subtracted this second spectrum from the first, and considered
the resultant spectrum to be a ``pure" $\beta$ spectrum.  This
method has one flaw, however: the spectrum obtained with the
aluminum plate includes some contribution from the bremmstrahlung
created by the $\beta$ particles as they stop.  Thus, we took the
same approach with the calculated Monte Carlo spectra: we made two
calculations for each source, one with an aluminum plate and one
without, took the difference between them and then compared that
difference spectrum with the ``pure" experimental $\beta$ spectrum.

\section{Geant4 Physics Model}

The Geant4 Simulation Toolkit includes a series of packages for the
simulation of electromagnetic interactions of particles with matter,
specialized for different particle types, energy range and specific
physics model. In our work, we considered only electrons, $\gamma$
rays and x rays, and used three different physics models for the
electromagnetic (EM) processes: the \textit{standard} EM package, the
\textit{low-energy} EM package and the \textit{Penelope} EM package. In all cases,
fluorescence emission, Rayleigh scattering and Auger interactions
were included in the EM physics model where appropriate. The
\textit{standard} EM package is based on an analytical
approach~\cite{Amako:2006,Burkhardt:2004,Ivantchenko:2005};
its effective energy range is nominally between 1~keV and 100~TeV but
it neglects atomic effects and is mainly used in high-energy physics
applications. The \textit{low-energy} package is optimized for our energy
region and extends the range of validity for electrons and photons
down to 250~eV~\cite{apostolakis:99,Chauvie:2004} and even below.
The \textit{Penelope} package is an alternative low-energy implementation; it
is a re-engineered version~\cite{Amako:2005} of the original
PENELOPE Monte Carlo code~\cite{Baro:1995,Sempau:2003}. Detailed
information on these packages, as well as on the design of the
Geant4 toolkit can be found in Ref.~\cite{agostinelli:03} and in the
Physical Reference Manual of Geant4~\cite{Colaboration:2005} and
references therein.

\begin{figure*}[]
  \begin{center}
  \includegraphics[angle=0,width=14cm, height=9.2cm]{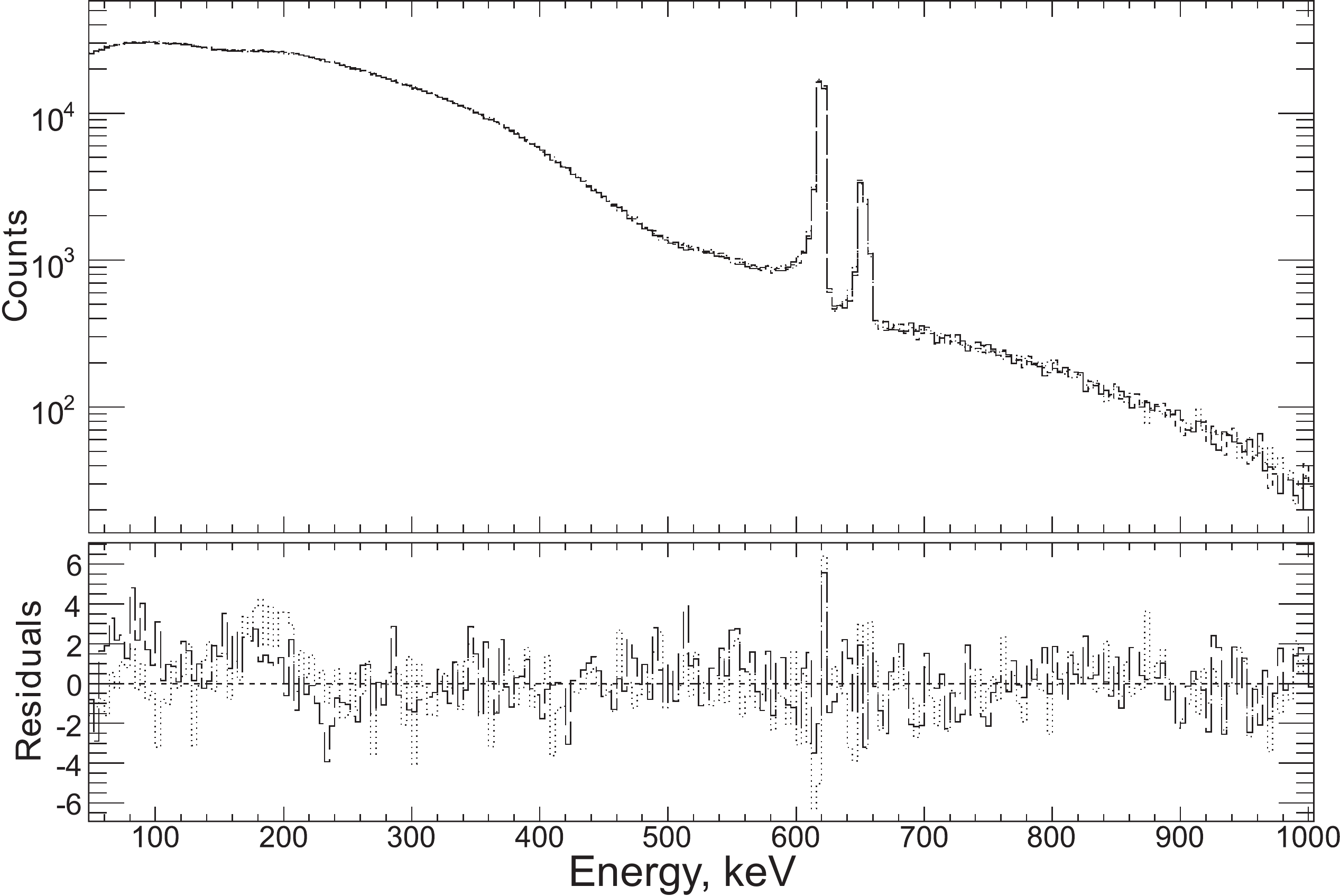}\\[-5pt]
  \caption{Upper panel: Energy deposition in the $\beta$-detector as generated by
  Geant4 Monte Carlo calculations for the decay of $^{137}$Cs. Three different EM
  physics models are used and all three are plotted although they are so similar
  that they cannot be distinguished. Lower panel: To clarify the differences among
  the results for the three EM physics models, the \textit{low-energy} EM package is used as
  our standard, and differences between each of the other two models and the \textit{low-energy}
  package are plotted as residuals in standard-deviation units.  The dashed line
  corresponds to the \textit{Penelope} EM package (compared with the \textit{low-energy} package);
  while the dotted line corresponds to the \textit{standard} EM package (also compared with
  the \textit{low-energy} package).}
  \label{fig:G4_EM_PhysModels}
  \end{center}
\end{figure*}

The validation of the Geant4 electromagnetic physical processes is
important in order to reach an adequate level of precision in
applications such as ours. Systematic and extensive validation is an
on-going process in the Geant4 collaboration and recently microscopic
quantities such as cross sections, angular/energy distributions,
attenuation coefficients, stopping powers and ranges have also been
examined in a systematic way~\cite{Cirrone:2003}, and their compatibility
with reference data from the U.S. National Institute of Standards and
Technologies (NIST) has been established~\cite{Amako:2005}.

\begin{table*}[t]
  \centering
  \caption{Composition of the different materials used in the Monte Carlo
  simulations performed in this work. The tabulated values correspond to the
  element mass fraction in each material, given in percentages.  Material
  densities are also given.}
\vspace{1mm}
\label{tab:materials}
  \begin{tabular*}{\textwidth}{@{\extracolsep{\fill}}c|rrrrrrrrr}
    \hline
    \hline
      \begin{tabular}{c}
          Chemical \\[-2mm]
          element \\
      \end{tabular}
      &  Air & Acrylic  & Mylar  & Havar  &
      \hspace{-10pt}
      \begin{tabular}{c}
          Stainless \\[-2mm]
          Steel \\
      \end{tabular}
        & PVC & Lucite & BC404 & Aluminum\\
    \hline
    H  &       &  0.71 &  4.20 &       &       &  4.84 &  8.07 &  8.45 &   \\[-1mm]
    C  &  0.01 &  8.52 & 62.50 &  0.04 &       & 38.44 & 59.97 & 91.55 &   \\[-1mm]
    N  & 75.53 & 90.77 &       &       &       &       &       &       &   \\[-1mm]
    O  & 23.18 &       & 33.30 &       &       &       & 31.96 &       &   \\[-1mm]
    Ar &  1.28 &       &       &       &       &       &       &       &   \\[-1mm]
    Be &       &       &       &  0.01 &       &       &       &       &   \\[-1mm]
    Cl &       &       &       &       &       & 56.73 &       &       &   \\[-1mm]
    Si &       &       &       &       &  1.00 &       &       &       &   \\[-1mm]
    Cr &       &       &       & 17.04 & 19.00 &       &       &       &   \\[-1mm]
    Ni &       &       &       & 12.50 & 10.00 &       &       &       &   \\[-1mm]
    Fe &       &       &       & 16.34 & 68.00 &       &       &       &   \\[-1mm]
    Co &       &       &       & 41.04 &       &       &       &       &   \\[-1mm]
    Mo &       &       &       &  3.14 &       &       &       &       &   \\[-1mm]
    Mn &       &       &       &  1.45 &  2.00 &       &       &       &   \\[-1mm]
    Al &       &       &       &       &       &       &       &       &  100.00 \\[-1mm]
    W  &       &       &       &  8.44 &       &       &       &       &   \\[1pt]
    \hline
    \begin{tabular}{c}
        Density \\[-2mm]
        g/{cm}$^{3}$ \\
    \end{tabular}
    & 0.0012 & 1.190 & 1.390 & 8.300 & 8.020 & 1.380 & 1.185 & 1.032 & 2.700 \\
    \hline
    \hline
  \end{tabular*}
\end{table*}

Here we have restricted ourselves to a comparison between Geant4
calculations and experiment for the energy deposited by electrons in
a thin plastic scintillator, where we have used the very simple laboratory
geometry already described so that the Monte Carlo geometry could reproduce
it exactly. As can be seen in the top panel of Figure~\ref{fig:G4_EM_PhysModels},
under these conditions the three physics models -- \textit{standard} EM, \textit{low-energy} EM
and \textit{Penelope} EM -- generate energy spectra that differ very little from
one to another.   The bottom panel of the figure shows the normalized
residuals between the first and second models, and between the second and
third ones: there are small but perceptable differences below 200 keV but
nothing significant above that energy.  We also compared the total $\beta$-efficiencies
obtained from the three EM physics models.  Including all energies between 50
and 1000~keV, the calculated Monte Carlo efficiencies were 15.16(3)\%,
15.18(3)\%, and 15.11(3)\%, respectively. If the low-energy threshhold
was increased to 75~keV the calculated efficiencies were 13.79(3)\%,
13.84(3)\% and 13.82(3)\%.  For both thresholds the three physics models
yielded statistically identical results.  In short, for our purposes we
find nothing in these results to choose between the three available physics models.

Even so, we chose to use the \textit{low-energy} EM model in all the Monte Carlo simulations
presented in the remaineder of this paper. Although it took considerably more computer
time per calculation than did the standard EM model, we considered that it was, in
principle, more appropriate to our energy region since it was specifically designed
for better performance at low energies.  A detailed
inter-comparison of results from the three physics models for the total efficiency of
the plastic $\beta$-detector as a function of $\beta$ energy will appear in a subsequent
publication~\cite{Golovko:2008}.

\section{Geant4 Geometry}

In defining the laboratory geometry in Geant4 we included all the
components of the detector assembly and source housing (see
Figs.~\ref{fig:betaDet}~and~\ref{RadioactiveSource2}, and
Table~\ref{tab:dimensions}), with everything placed in air. Special
care was taken to include all components of the various
materials, with the natural isotopic abundances for each
element properly accounted for. Table~\ref{tab:materials} lists
the composition and density of all materials used in the Geant4
calculations.

\begin{figure}[!t]
  \includegraphics[width=\columnwidth, height=12.5cm]{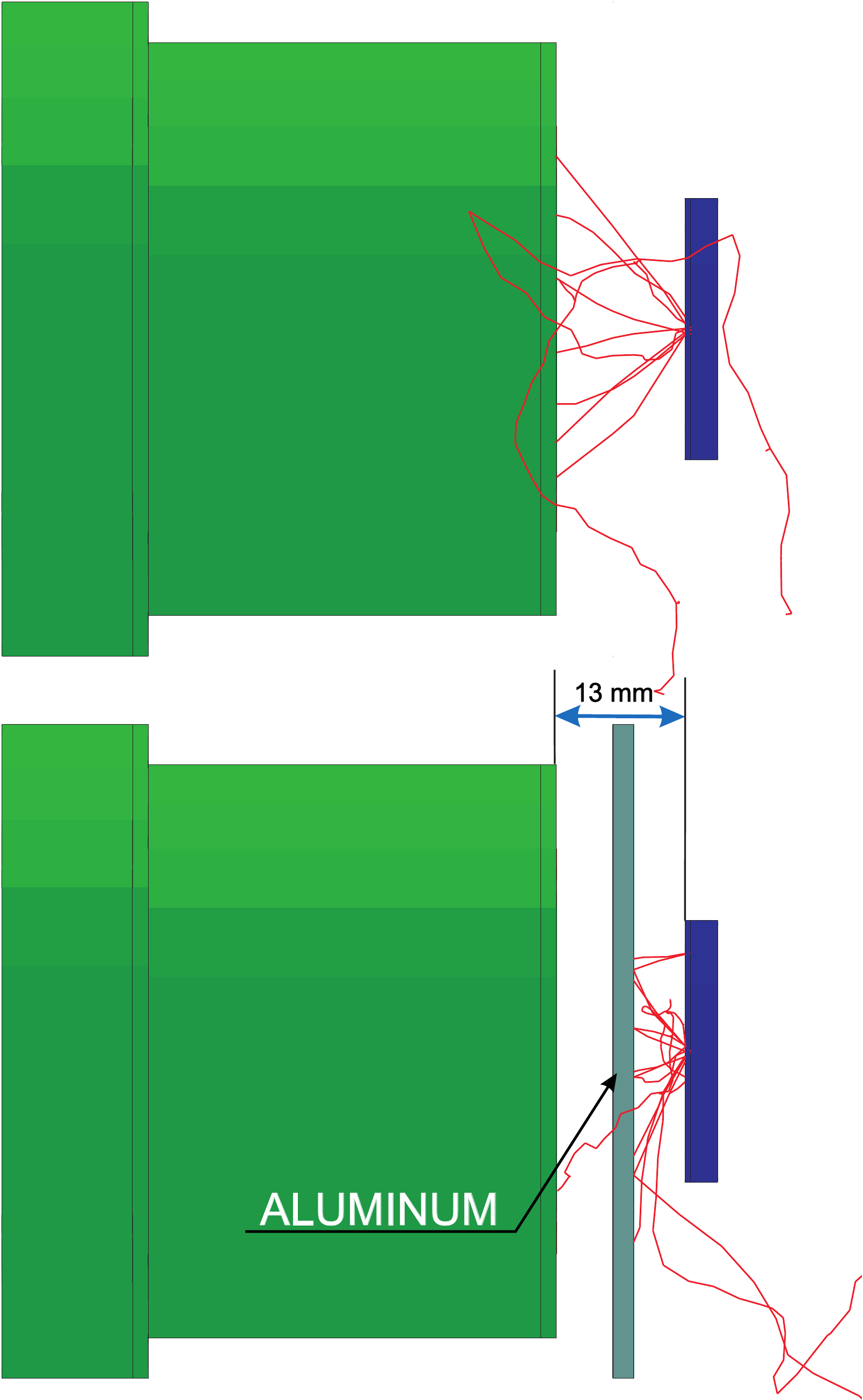}\\[-20pt]
  \caption{Schematic view of the geometry arrangements for
  Monte Carlo simulations, showing sample electron trajectories.
  The upper panel shows trajectories when there is only air between
  the source holder (right) and the detector housing (left). The
  lower panel shows them when a 2-mm-thick aluminum plate is
  introduced.}
  \label{fig:DetAlAndAir}
\end{figure}

As explained in Section~\ref{Det}, two Monte Carlo simulations were
performed for each source, one with a 2-mm-thick plate of aluminum
located between the source and the detector, and one without it.  The
spectrum we plot for each source is the difference between these two
spectra. In defining the geometry in Geant4 we incorporated a plate
and chose its material to be either aluminum or air, depending on the
desired effect. Fig.~\ref{fig:DetAlAndAir} shows the tracks of electrons
for both types of simulations.

\section{Geant4 Parameter Control}

Geant4 allows the user to define different regions in the
experimental setup, and to set a different particle-production
threshold in each one~\cite{Allison:2006}. This capability allows
for simulation accuracy and speed optimization according to the
needs of a particular experiment. We defined three regions: the
radioactive source, the thin plastic scintillator and everything
else.  In the first two, the threshold for producing secondary
particles was kept very low in order to simulate all physics
interactions as well as possible; in the third, a much higher
production threshold was chosen. This approach considerably reduced
the computing time without compromising the results. A test run with
low thresholds in all regions did not reveal any significant
differences from the sped-up version with different thresholds in
different regions.

It was pointed out by Kraev~\cite{Kraev:2006} that, to get good
agreement with experiment, it is important to choose two parameters
particularly carefully in the \textit{low-energy} EM physics model of Geant4:
the cut for secondaries (\textit{CFS}) value, which determines the
production threshold for secondary particles, and the
$f_r$-parameter, which limits the step size for tracking
$\beta$-particles at a material boundary.  In this work, we used
\textit{CFS} = 10~$\mu$m, which is recommended as an acceptable
compromise between a ``good" description of the scattering processes
and a reasonable computation time. For the $f_r$-parameter we used
0.02 as recommended in~\cite{Kraev:2006}; this
is actually the default value for the Geant4 version~4.9.0.

\section{Comparison with Experiment}

To simulate the decay of a radioactive nuclide with Geant4, it is
possible to define each $\gamma$ transition, internal-conversion
line and $\beta$-decay spectrum individually and require Geant4 to
transport all particles through the specified materials and
determine the spectrum in the scintillator.  However, the code also
offers a radioactive-decay module, which generates all the decay
components radiated from a specified source using information
extracted from the Evaluated Nuclear Structure Data File
(ENSDF)~\cite{Tuli:2001}.

\begin{figure}[!b]
  \includegraphics[width=\columnwidth]{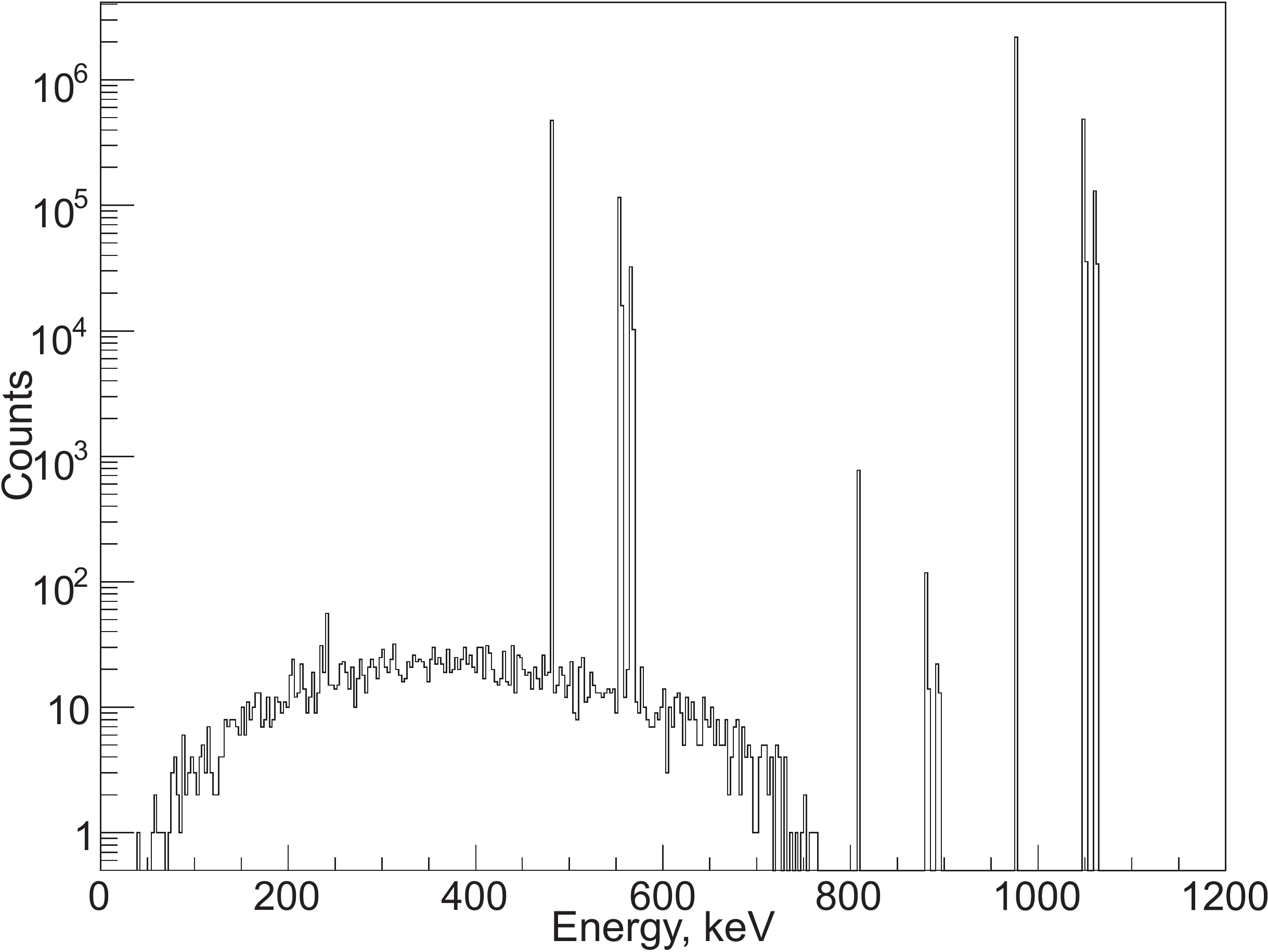}\\[-20pt] %
  \caption{Decay spectrum for $^{207}$Bi generated by Geant4 with
  its internal radioactive-decay module activated.  Only electrons
  are shown.}
  \label{fig:betaEmit}
\end{figure}

To generate the Monte Carlo emission spectra we began by programming
Geant4 based on the radioactive-decay module. The primary electron
spectrum emitted from $^{207}$Bi generated with the
radioactive-decay module is shown in Fig.~\ref{fig:betaEmit}. When
repeating this procedure for $^{133}$Ba, to our surprise we found
that the electron emission spectrum produced by the
radioactive-decay module of Geant4 was simply not correct, yielding
relative conversion-electron intensities in significant disagreement
with ENSDF data.  The emission spectrum from $^{137}$Cs also turned
out to be incorrect, but here the main problem was more subtle:
there are two $\beta$-decay branches from $^{137}$Cs, which are both
treated by Geant4 as allowed.  In fact both transitions are
forbidden, with shape-correction factors that have been determined
by Behrens and Christmas~\cite{Behrens:1983} from experimental data.
In addition, the radioactive-decay module gives the incorrect
intensity for one of the conversion electron lines of $^{137}$Cs
($I_{655.7~\rm{keV}}$=1.10\% instead of 1.39\%).  In both these
decays -- of $^{133}$Ba and $^{137}$Cs -- we bypassed the
radioactive-decay module and inserted each decay mode and transition
individually, with the correct intensities for the conversion
electrons and the correct shape for the forbidden $\beta$
transitions. For these we used the General Particle Source module
available in the Geant4~\cite{GPSweb}, which allows the user to
define standard energy, angle and space distributions of the primary
particle.

\begin{figure}[t]
  \includegraphics[angle=0,width=\columnwidth]{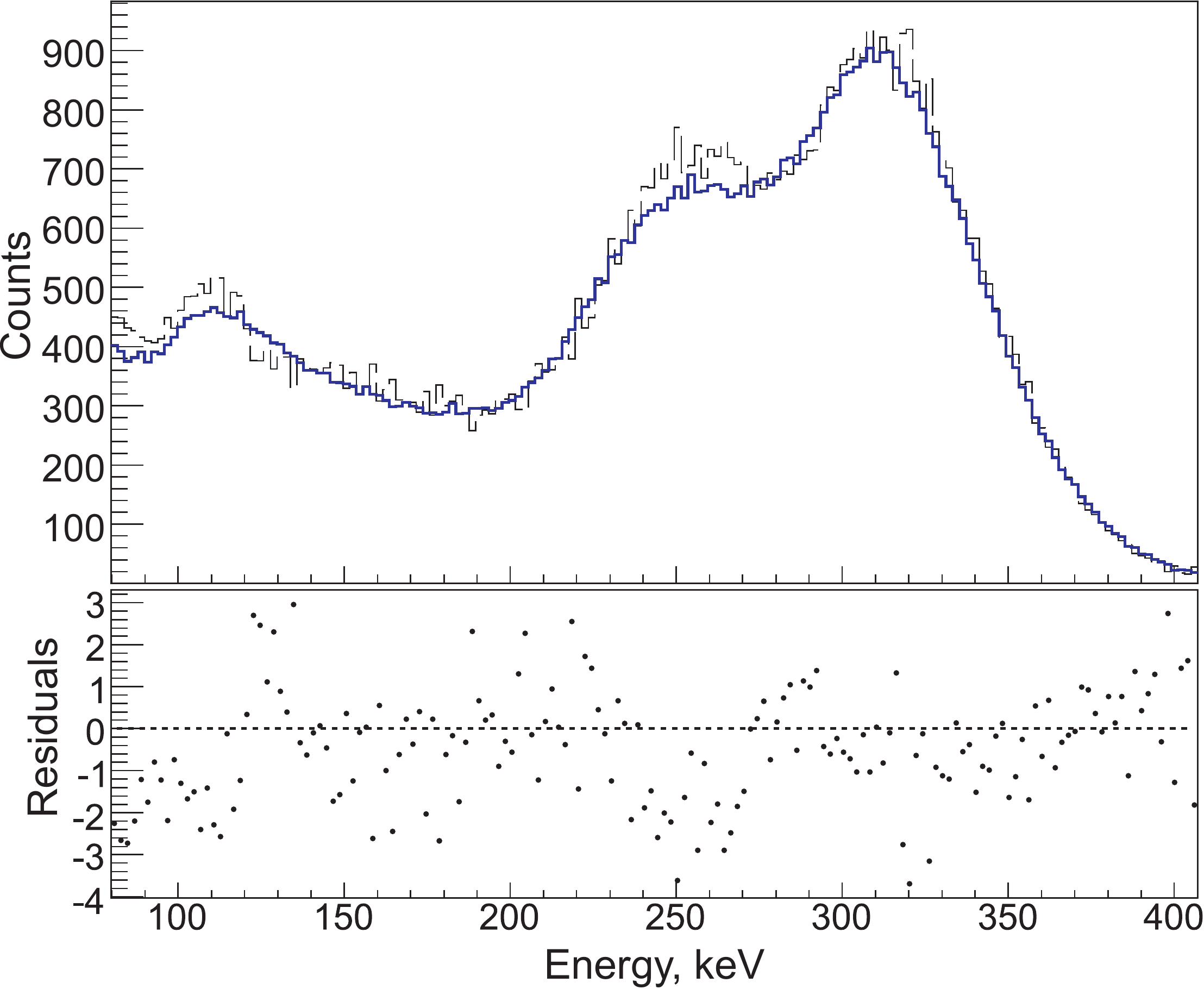}\\[-20pt]
  \caption{In the top panel, the measured spectrum (thick solid line) for the decay
   of $^{133}$Ba is compared with the Geant4-simulated result (thin dashed line).
   The \textit{low-energy} EM package was used.  Residuals in standard-deviation units are plotted
   in the lower panel.  The reduced $\chi^2$ in the energy range 80 -- 406~keV is 0.4.}
  \label{fig:Ba133}
\end{figure}

Based on a primary spectrum thus generated for each source, the
Monte Carlo code then determined the total energy deposited in the
scintillator.  However, before this result could be compared with
the experimental spectrum, it was necessary to add the effects of
statistical fluctuations introduced by the processes of light
production and transmission, as well as photomultiplication and
electronic pulse analysis. For this purpose, we looked to a
published study of the response of a plastic scintillator to
mono-energetic beams of positrons and electrons
\cite{Clifford:1984}, which tabulated
the width of the full-energy Gaussian peak as a function of energy
between 0.8 and 3.8~MeV. Since we also needed to deal with energies
lower than that, we took the width to be linearly dependent on
energies below 0.8~MeV.

Our procedure was to take the scintillator spectrum produced by
Geant4 and process it by a randomization algorithm written in
C${++}$ in the ROOT \cite{Brun:1997} analysis framework.  In
essence, this process spread the number of counts in each energy bin
into a Gaussian distribution centered at the original energy and
with a width, $\sigma$, taken or extrapolated from
Ref.~\cite{Clifford:1984}.  The results could then be compared
directly with the measured spectra.

The measured spectra for all three sources were taken under exactly the
same conditions. A description of the data acquisition system used
in these measurements is given in ref.~\cite{iacob:015501} and
references therein. The amplifier gain, photomultiplier high voltage, and
low-energy electronic threshold remained unchanged for all three sources. The
threshold was chosen to be identical to the setting used during our
branching-ratio measurements of superallowed Fermi $\beta$-decays. (This hardware
threshhold was higher than the software threshhold.) This ensured
that the results of our comparisons between source measurements and
Monte Carlo simulations could be readily applied to our accelerator-based
measurements.

\begin{figure}[t]
  \includegraphics[angle=0,width=\columnwidth]{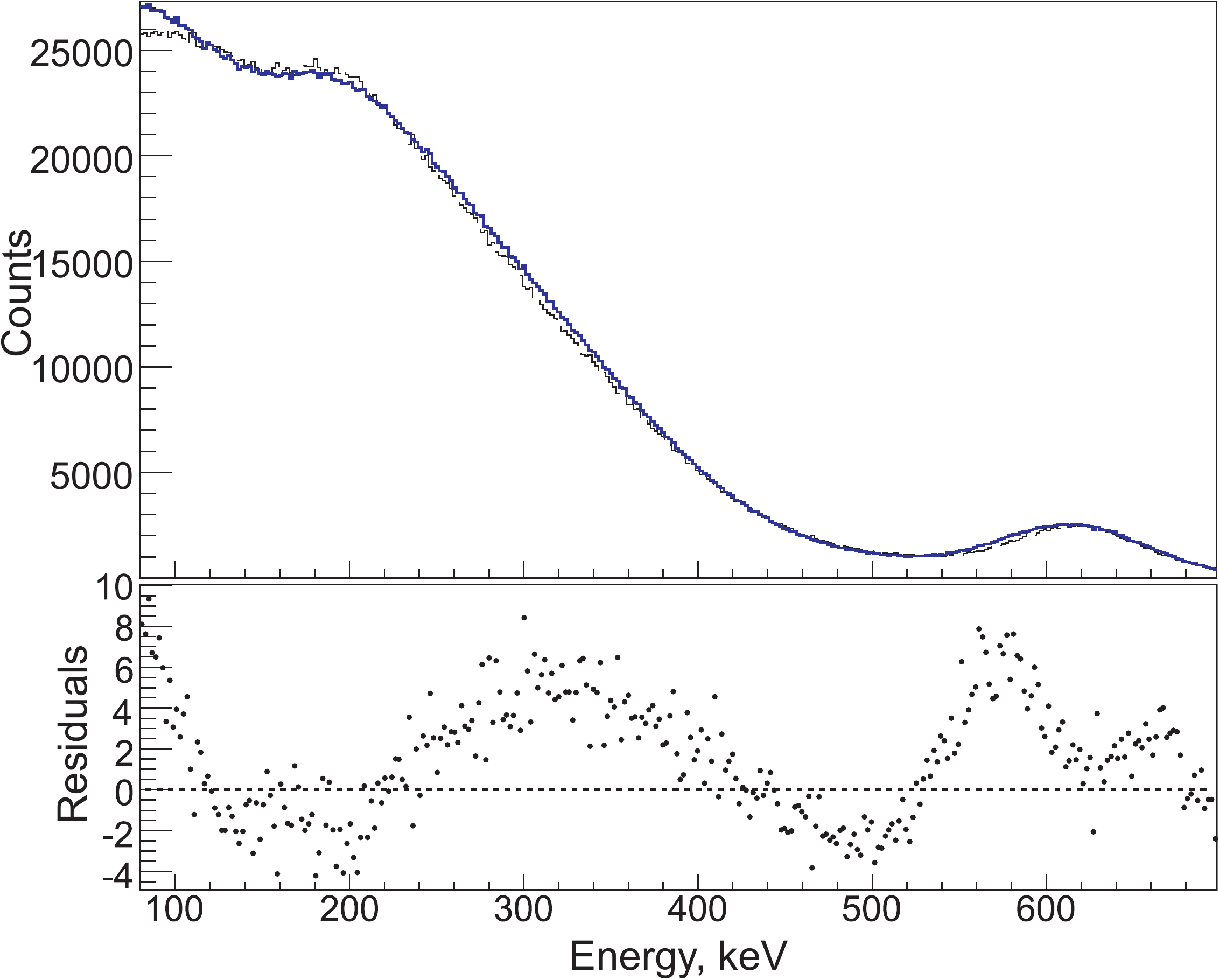}\\[-20pt]
  \caption{In the top panel, the measured spectrum (thick solid line) for the decay
   of $^{137}$Cs is compared with the Geant4-simulated result (thin dashed line).
   The \textit{low-energy} EM package was used.  Residuals in standard-deviation units are plotted
   in the lower panel.  The reduced $\chi^2$ in the energy range 80 -- 697~keV is 4.0.}
  \label{fig:Cs137}
\end{figure}

\begin{figure}[t]
  \includegraphics[angle=0,width=\columnwidth]{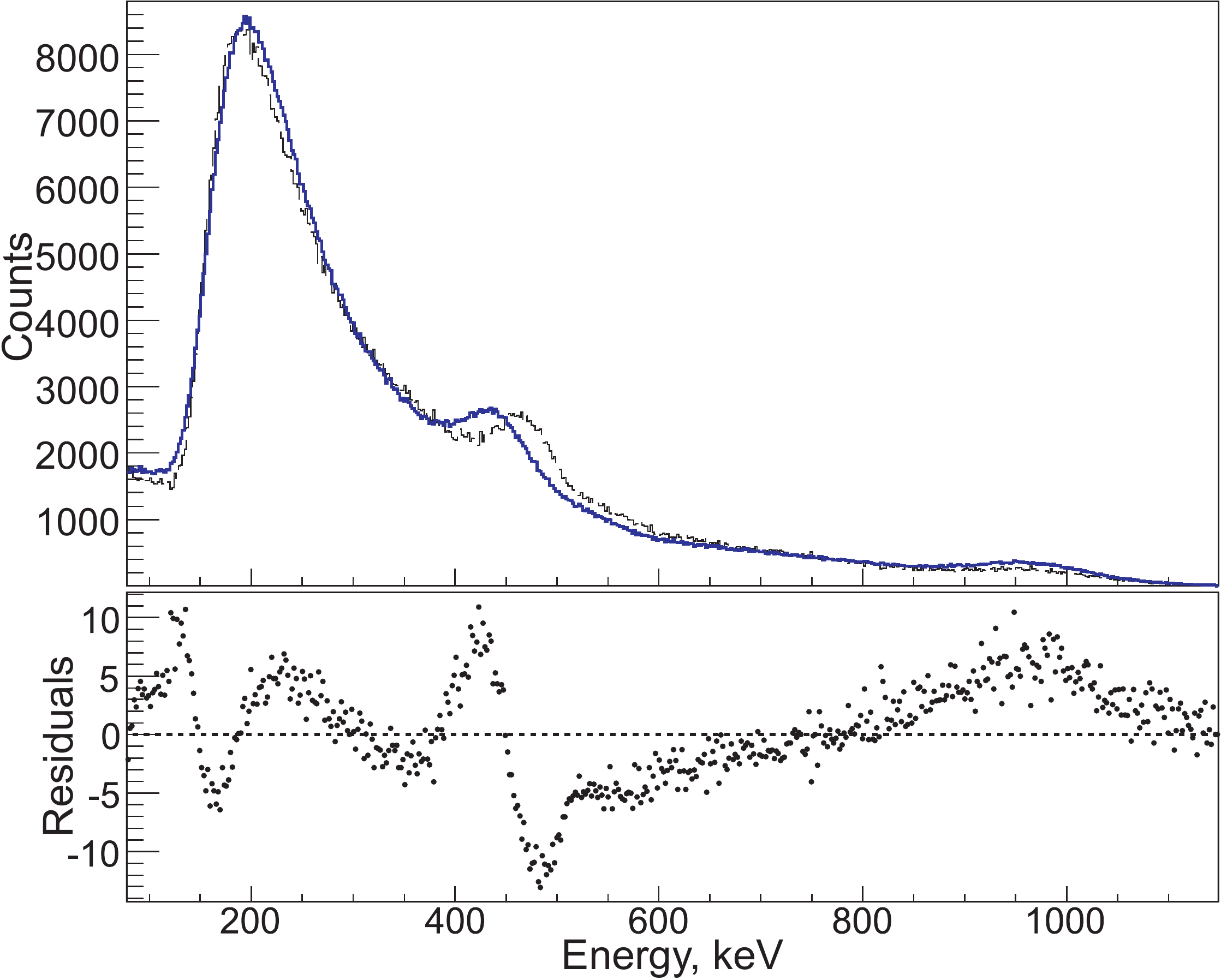}\\[-20pt]
  \caption{In the top panel, the measured spectrum (thick solid line) for the decay
   of $^{207}$Bi is compared with the Geant4-simulated result (thin dashed line).
   The \textit{low-energy} EM package was used.  Residuals in standard-deviation units are plotted
   in the lower panel.  The reduced $\chi^2$ in the energy range 80 -- 1143~keV is 8.9.}
  \label{fig:Bi207}
\end{figure}

In comparing our measured spectra with Monte Carlo results, we the slope of
the energy calibration ({\it i.e.} the energy per channel) and its offset
(({\it i.e.} the zero-energy channel) corresponding to each measured
spectrum as adjustable fit parameters, which were used to
optimize the agreement with the Monte Carlo spectrum.  We wrote our own C++ ROOT
program to accomplish this purpose.  The resulting comparisons for our
three sources, as well as the normalized residuals, for $^{133}$Ba,
$^{137}$Cs and $^{207}$Bi, appear in Figures~\ref{fig:Ba133},
\ref{fig:Cs137} and \ref{fig:Bi207} respectively.

The agreement between the Geant4 Monte Carlo simulations and experiment
is good for all three sources considered in this work, $^{133}$Ba,
$^{137}$Cs and $^{207}$Bi, although for the latter one the normalized
$\chi^2$ of 8.9 is less impressive than for other two. In that case, the
energy range, which extended from 80 to 1143~keV, was much greater than
for the other too. This may be partly responsible for the higher $\chi^2$
but another possibility is that our simple linear extrapolation of the
results in Ref.~\cite{Clifford:1984} does not fully describe our system's
response function at low energies.  If in fact the peak resolution were
somewhat worse than this extrapolation indicates -- a not unreasonable
possibility -- then the agreement with experiment would be considerably
improved.  It is also worth noting that the slight non-allignment in
peak positions, which is evident in all three spectra, can be
explained by possible small non-linearities in the experimental
energy response.

Since the amplifier gain and photomultiplier high voltage were kept
the same for all three measurements, the fitted slopes and offsets obtained
for all three spectra should have been very nearly the same.  In fact they
were, but, as a quantitative measure of consistency, we took the
average values for the slope and offset, and again evaluated the normalized
$\chi^2$ for the comparison between experiment and simulation.  The new
values for the normalized $\chi^2$ were, of course, somewhat increased,
being 3.6, 9.5, and 16.3 for $^{133}$Ba, $^{207}$Bi, and $^{137}$Cs,
respectively, but the agreement is still quite satisfactory.  Most crucially
for our purposes, the low-energy thresholds in all three cases were in close
agreement, with 80$\pm$3~keV being the common value.

As stated in the introduction, the result of the simulation that matters
most to us is how well it reproduces the fraction of the total $\beta$
spectrum that lies above some low-energy threshold, typically around 80~keV.  The
spectral details at higher energy are only important to us to the extent
that they change the fraction of counts recorded above threshold. The fact
that the fits performed with the low-energy threshold as a free parameter
give the same result for all three isotopes, serves as important assurance
that the fitted threshold value is consistent with the actual one.

Although the activity of the radioactive sources that we used for
this work are nominally 1~$\mu$Ci (37-kBq), the accuracy of this value
was only quoted to and approximate $\pm$15\% by the supplier.  So that
we could get a more precise value for our $\beta$-detector efficiency,
we made our own measurement of the $^{207}$Bi source activity using a
well-calibrated HPGe $\gamma$-detector~\cite{Hardy:2002} to detect the
known $\gamma$ rays from the decay.  In this way we established the source
activity to be 1.31(1)~$\mu$Ci at 16 January, 2008.  Now, knowing the activity
of the source as well as the low-energy detection threshold already obtained from
our fit, we could deduce from our experimental data the absolute efficiency
of the $\beta$-detector to be 3.48(2)\% at the distance of 13.2(1)~mm. With
exactly this geometry, the Geant4 simulation yielded an absolute efficiency
of 3.50(1)\%, in excellent agreement with experiment.

\section{Conclusion}

The electron spectra we have obtained from Monte Carlo simulations
are generally in good agreement with experimental data. All
identifiable features that are present in the experimental spectra
are reproduced in the simulated ones, and in most cases their
relative intensities agree as well. Furthermore, the threshold energy
and absolute efficiency are well reproduced. Our results clearly demonstrate
that Geant4 version~4.9.0 can be used effectively to simulate the
$\beta$-spectra as measured by our thin plastic scintillator
detector.

In particular for our application, where we need only rely on the
simulation to determine the energy dependence of the $\beta$-detector's
total efficiency with a low-energy threshold at $\sim$80~keV, it is clear
that Geant4 will provide the precision we require.

However, we have also shown that the radioactive-decay module
included in Geant4 to simulate the initial radiation from a
radioactive source should only be used with care: it is essential to
check its output carefully.  In particular, the intensities of
conversion-electron lines produced by this module were sometimes
found to be incorrect and, if any forbidden $\beta$-decay branches
are involved, their spectrum shapes may not be correctly generated.


\begin{thebibliography}{27}

\bibitem{hardy:055501}
J.~C. Hardy and I.~S. Towner, Phys. Rev. C 71 (2005) 055501.

\bibitem{Towner:025501}
I.~S. Towner and J.~C. Hardy, Phys. Rev. C 77, (2008) 025501.

\bibitem{agostinelli:03}
S.~Agostinelliae {\it et al.}, Nucl. Instr. and Meth. A 506 (2003)
250, and Geant4 Home Page, \url{http://geant4.cern.ch/}.

\bibitem{iacob:015501}
V.~E. Iacob {\it et al.}, Phys. Rev. C 74 (2006) 015501.

\bibitem{Hardy2003}
J.~C. Hardy {\it et al.}, Phys. Rev. Lett. 91 (2003) 082501.

\bibitem{Helmer:2003}
R.~G. Helmer {\it et al.}, Nucl. Instr. and Meth. A 511 (2003) 360.

\bibitem{apostolakis:99}
J.~Apostolakis {\it et al.}, CERN-OPEN-034 (1999).

\bibitem{Chauvie:2004}
S.~Chauvie {\it et al.}, in: Nuclear Science Symposium Conference
Record, 2004 IEEE, Vol.~3, (2004) 1881.

\bibitem{Allison:2006}
J.~Allison {\it et al.}, IEEE Trans. Nucl. Sci. 53 (2006) 270.

\bibitem{LaserSensor_web}
AccuRange 600$^{\rm{TM}}$ Laser Displacement Sensor Home Page,
\url{http://www.acuitylaser.com/AR600/sensor-technical-data.shtml}.

\bibitem{Amako:2006}
K.~Amako {\it et al.}, Nucl. Phys. B, Proc. Suppl. 150 (2006) 44.

\bibitem{Burkhardt:2004}
H.~Burkhardt {\it et al.}, in: Nuclear Science Symposium Conference
Record, 2004 IEEE, Vol.~3, (2004) 1907.

\bibitem{Ivantchenko:2005}
V.~Ivanchenko {\it et al.}, in: Proceedings of the CHEP'04, CERN
2005-002 Vol.~1, (2005) 207.

\bibitem{Amako:2005}
K.~Amako {\it et al.}, IEEE Trans. Nucl. Sci. 52 (2005) 910.

\bibitem{Baro:1995}
J.~Baro {\it et al.}, Nucl. Instr. and Meth. B 100 (1995) 31.

\bibitem{Sempau:2003}
J.~Sempau {\it et al.}, Nucl. Instr. and Meth. B 207 (2003) 107.

\bibitem{Colaboration:2005}
Geant4~Colaboration, Physics Reference Manual, CERN (2007),
\url{http://geant4.cern.ch/support/userdocuments.shtml}.

\bibitem{Cirrone:2003}
G.~Cirrone {\it et al.}, in: Nuclear Science Symposium Conference
Record, 2003 IEEE, Vol.~1, (2003) 482.

\bibitem{Golovko:2008}
V.V.~Golovko {\it et al.}, to be published.

\bibitem{Kraev:2006}
I.~S. Kraev, Ph.D. thesis, 
  KU Leuven (2006).

\bibitem{Tuli:2001}
J.~Tuli, 
 Brookhaven National Lab report, BNL-NCS-51655-02-Rev (2001).

\bibitem{Behrens:1983}
H.~Behrens and P.~Christmas, Nucl. Phys. A399 (1983) 1310.

\bibitem{GPSweb}
General Particle Source Home Page,
\url{http://reat.space.qinetiq.com/gps/}.

\bibitem{Clifford:1984}
E.~T.~H. Clifford {\it et al.}, Nucl. Instr. and Meth. 224 (1984)
440.

\bibitem{Brun:1997}
R.~Brun and F.~Rademakers, Nucl. Instr. and Meth. A 389 (1997) 81.

\bibitem{Ortec_web}
Ortec TRUMP Home Page, \url{http://www.ortec-online.com/trump.htm}.

\bibitem{Hardy:2002}
J.~C. Hardy {\it et al.}, Appl. Rad. and Isotopes 56(1--2) (2002)
65--69.

\end{thebibliography}
\end{document}